 \documentstyle[prb,twocolumn,eqsecnum,aps]{revtex}
\begin{document}
\draft
\title{Curvature effects on the surface thickness and tension at
the free interface of $^4$He systems}
\author{Leszek Szybisz$^{1,2,}$\cite{szybco} and Ignacio
Urrutia$^{1,}$\cite{urrucic}}
\address{$^1$Laboratorio TANDAR, Departamento de F\'{\i}sica,
Comisi\'on Nacional de Energ\'{\i}a At\'omica,\\
Av. del Libertador 8250, RA-1429 Buenos Aires, Argentina}
\address{$^2$Departamento de F\'{\i}sica, Facultad de
Ciencias Exactas y Naturales,\\
Universidad de Buenos Aires,
Ciudad Universitaria, RA-1428 Buenos Aires, Argentina}
\date{\today}
\maketitle
\begin{abstract}
The thickness $W$ and the surface energy $\sigma_A$ at the free
interface of superfluid $^4$He are studied. Results of calculations
carried out by using density functionals for cylindrical and
spherical systems are presented in a unified way, including a
comparison with the behavior of planar slabs. It is found that
for large species $W$ is independent of the geometry. The obtained
values of $W$ are compared with prior theoretical results and
experimental data. Experimental data favor results evaluated by
adopting finite range approaches. The behavior of $\sigma_A$ and $W
\sigma_A$ exhibit overshoots similar to that found previously for
the central density, the trend of these observables towards their
asymptotic values is examined.


\end{abstract}

\section{Introduction}
\label{sec:introduce}

The understanding of the density profiles in the surface region of
a quantum fluid like $^4$He has long been considered a very
important basic problem.\cite{edda,os,sill} At the liquid-vapor
interface, the density profile of $^4$He changes continuously from
liquid density $\rho_\ell$ to vapor density $\rho_v$ over a
distance of some {\aa}ngstr\"oms. In the case of a liquid-vacuum
interface at $T=0$~K, $\rho_\ell$ falls monotonically to $\rho_v =
0$. The width $W$ of a surface is defined as the distance in which
the density decreases from $0.9\rho_\ell$ to $0.1\rho_\ell$.
A glance at recent literature reveals several works addressing the
question of the thickness of the free interface,\cite{dali,eu,%
gall,penal} indicating the continuous interest in this area of
theoretical and experimental research. The surface tension at the
free interface has been also investigated for a long time.\cite{%
edda,eu,gest,isis,raw,panda,vale} A list of results for the surface
thickness and tension determined up to the middle of 1987 is given
by Osborne in Table 1 of Ref. \onlinecite{os}.

In the systematic study of free planar $^4$He films at $T=0$~K we
have, among other issues, discussed features of the surface
thickness.\cite{eu} There, our own results for $W$ evaluated by
using several density functional (DF) approaches are compared with
values obtained from Monte Carlo simulations by Vall\'es and
Schmidt\cite{vale} and experimental data of Lurio {\it et al.}\cite{sill}
From Fig. 4 in Ref. \onlinecite{eu} one can realize that the size of the
experimental error bar was to large for disregarding any of the
applied DF approaches. After that paper, new theoretical and
experimental results for this quantity have appeared. Evaluations of
$W$ for free planar slabs and droplets by utilizing a variational
Monte Carlo with shadow wave functions with a glue term
(glue-SWF) were published by Galli and Reatto.\cite{gall} On the
other hand, a novel measurement of $W$ by using x--ray reflectivity
was reported by Penanen {\it et al.}\cite{penal} These new data
supersede that previously obtained at the same laboratory.\cite{sill}
To complete the survey of investigations about the surface thickness
of superfluid helium droplets, one should mention the calculation
of Stringari and Treiner\cite{sat1} and the comprehensive
experimental and theoretical study of Harms, Toennies, and
Dalfovo.\cite{dali}

The foregone summary indicates that there is an important piece of
information about the free surface thickness of planar and
spherical superfluid $^4$He systems. Although in recent years
there is a renewal of interest for examining cylindrical species,
\cite{stucco,coal3,gat,cyan,less,hall1,ur} hitherto, there is no
study undertaking the problem of exploring features of $W$ in the
case of such geometry.

On the other hand, we have explored the evolution of the surface
energy at the liquid-vacuum interface of planar slabs as a function of
their size in Ref. \onlinecite{eu}. However, as far as we know, there is
no work devoted to study this property in the case of systems with
curved geometries. Therefore, such an analysis becomes an
interesting problem by its own right.

In view of the situation described in the previous paragraphs, the
aim of this work is to study systematically the interface thickness
and the surface tension of liquid $^4$He with cylindrical and
spherical shapes making a connection to the case of planar slabs.
Although the spherical systems are energetically favored against
free cylinders and slabs an analysis of the overall picture
presents instructive features. The width at the liquid-vacuum
interface of free systems is compared with experimental data and
with theoretical results obtained for stable planar films of $^4$He
adsorbed onto the lightest alkali metals.
The theoretical tools are outlined in Sec. \ref{sec:theory}. In
Sec. \ref{sec:large} we search for the size at which the systems
reach an asymptotic global behavior. The discussion of the pattern
exhibited by the width and tension at the surface is done in Sec.
\ref{sec:study}, where the results are presented in a unified way
allowing a direct comparison of data obtained for different
geometries. Section \ref{sec:conclude} is devoted to a summary.

\section{Theoretical Framework}
\label{sec:theory}

The calculations performed in the present work were carried out by
using density functional (DF) approaches, which have proven to be
successful tools for treating this kind of quantum many-body
problems. In such a theory the ground-state energy, $E_{\rm gs}$,
of an interacting $N$-body system of $^4$He atoms, confined by an
adsorbate-substrate potential $U_{\rm sub} ({\bf r})$, may be
written as
\begin{eqnarray}
E_{\rm gs}~&&= \int { d{\bf r}\,\rho({\bf r})\,{\cal{H}}[\rho,{\bf
\nabla}\rho]}
+ \int { d{\bf r}\,\rho({\bf r})\,U_{\rm sub}({\bf r}) }
\nonumber\\
&&= -{\hbar^2\over2 m} \int { d{\bf r} \,
\sqrt{\rho({\bf r})} {\bf \nabla}^2\,\sqrt{\rho({\bf r})} }
+ \int { d{\bf r}\,\rho({\bf r})\,e_{sc}({\bf r}) }
\nonumber\\
&&~~~+ \int { d{\bf r}\,\rho({\bf r})\,U_{\rm sub}({\bf r}) } \;,
\label{Ene0}
\end{eqnarray}
where $\rho({\bf r})$ is the one-body density. The first term on
the right-hand side is the quantum kinetic energy of the helium
particles of mass $m$. The second term represents the interaction
between the particles of the system, where $e_{sc}({\bf r})$ is
the self-correlation energy per particle depending on the DF
approach. The last term is the interaction with the external field.

The density profile $\rho({\bf r})$ is determined from the
Euler-Lagrange (EL) equation derived from the condition
\begin{equation}
\frac{\delta \Omega}{\delta \rho({\bf r})}
= \frac{\delta \{\,E_{\rm gs}[\rho,{\bf \nabla}\rho] - \mu\,N\,\}}
{\delta \rho({\bf r})} =  0 \;. \label{vary0}
\end{equation}
Here $\mu$ is the chemical potential, $N$ the number of particles
\begin{equation}
N = \int { d{\bf r} \, \rho({\bf r}) } \;, \label{np}
\end{equation}
and $\Omega$ the grand thermodynamic potential.
The variation of (\ref{vary0}) leads to a Hartree like equation for
the square root of the one-body density 
\begin{equation}
\biggr[ -\frac{\hbar^2}{2 m} {\bf \nabla}^2 + V_H({\bf r})
+ U_{\rm sub}({\bf r}) \biggr] \,\sqrt{\rho({\bf r})} = \mu \,
\sqrt{\rho({\bf r})} \;, \label{hair0}
\end{equation}
which also determines $\mu$. Here $V_H({\bf r})$ is a Hartree
mean-field potential given by the first functional derivative of
the total correlation energy $E_{sc}[\rho]$
\begin{eqnarray}
V_H({\bf r})~&&= \frac{\delta E_{sc}[\rho,{\bf \nabla}\rho]}
{\delta \rho({\bf r})}
= \frac{\delta}{\delta \rho({\bf r})} \, \int d{\bf r}\prime \,
\rho({\bf r}\prime)\,e_{sc}({\bf r}\prime) \nonumber\\
&&= \biggr[ \frac{\partial}{\partial \rho({\bf r})} - {\bf \nabla}
\frac{\partial}{\partial {\bf \nabla}\rho({\bf r})} \biggr] \,
\int d{\bf r}\prime \,\rho({\bf r}\prime)\,e_{sc}({\bf r}\prime)
\;. \label{harp}
\end{eqnarray}
Two different DF were used, namely, the Skyrme type ``zero-range''
version suggested in Refs. \onlinecite{sat1} and \onlinecite{sat2},
and the nonlocal density functional (NLDF) proposed in Ref.\
\onlinecite{dhpt}. For each DF an expression for $V_H(r)$ should be
derived.

\subsection{Zero-range density functional}
\label{sec:zero}

The simplest DF successfully employed to interpret properties of
$^4$He systems is a zero-range correlation proposed by Stringari
and Treiner.\cite{sat1,sat2} It has been inspired in Skyrme type
functionals extensively used to describe properties of atomic
nuclei.\cite{ring,string} The explicit form of this correlation
energy per particle is
\begin{equation}
e^{\rm Sky}_{sc}({\bf r}) = \frac{b_4}{2}\,\rho({\bf r})
+ \frac{c_4}{2}\,\rho^{\gamma_4+1}({\bf r}) + d_4\,
\frac{1}{\rho({\bf r})}\,\mid \nabla \rho({\bf r}) \mid^2 \;.
\label{sky}
\end{equation}
The phenomenological parameters $b_4$, $c_4$, $\gamma_4$, and $d_4$
have been fixed in Ref.\ \onlinecite{sat1} so as to reproduce the
known observables of the bulk liquid at equilibrium. The data of
these saturation quantities (where $P=0$), {\it i.e.}, the
equilibrium density $\rho_0$, the minimum energy per particle
$e_B$, the compressibility $\cal{K}$, and the surface tension
$\sigma_\infty$ of a semi-infinite $^4$He system are listed in
Table\ \ref{table1}. Experimental values are correctly reproduced
by the set
\begin{eqnarray}
\begin{array}{l}
b_4 = -8.888\,10 \times 10^2 \, {\rm K\,\AA^3} \;, \\
c_4 = 1.045\,54 \times 10^7 \, {\rm K\,\AA^{3(\gamma_4+1)}} \;, \\
\gamma_4 = 2.8 \;, \\
d_4 = 2.383 \times 10^3 \, {\rm K\,\AA^5} \;.
\end{array} \label{values}
\end{eqnarray}

\subsection{Orsay-Paris nonlocal density functional}
\label{sec:trento}

Currently, in the literature one may find a few NLDF approaches.
The Orsay-Paris (OP) version (OP-NLDF) developed by Dupont-Roc {\it
et al.}\cite{dhpt} treats correctly the long-range part of the
helium-helium interaction and provides a reasonable description of
correlations. The most elaborated version has been formulated by
the Orsay-Trento (OT) collaboration.\cite{dale} However, the
OP-NLDF is sufficiently good to reproduce properties of non-layered
samples like free or weakly confined systems. This functional reads
\begin{equation}
e^{\rm OP}_{sc}({\bf r}) = \frac{1}{2}\,\int { d{\bf r}\prime \,
\rho({\bf r}\prime)\,V^{\rm OP}_l(\mid {\bf r} - {\bf r}\prime
\mid) } + \frac{c_4}{2}  \, [\,\bar\rho({\bf r})\,]^{\gamma_4+1}
\;. \label{dupe2}
\end{equation}
In this case the two-body interaction, $V_l^{\rm OP}(\mid {\bf r} -
{\bf r}\prime \mid)$, was taken as the $^4$He-$^4$He Lennard-Jones
(LJ) potential screened in a simple way at distances shorter than a
characteristic distance $h_{\rm OP}$
\begin{eqnarray}
\: V^{\rm OP}_l(r) = \left\{
\begin{array}{lll}
4 \epsilon_{\rm LJ} \biggr[\left(\frac{\sigma_{\rm LJ}}{r}
\right)^{12} - \left(\frac{\sigma_{\rm LJ}}{r}\right)^6\biggr] &
{\rm if} & r \, \geq h_{\rm OP} \;, \\
&& \\
V^{\rm OP}_l(h_{\rm OP})\,\left(\frac{r}{h_{\rm OP}}\right)^{4} 
& {\rm if} & r \, < h_{\rm OP} \;, \\
\end{array} \right. \label{screen}
\end{eqnarray}
with the standard de Boer and Michels parameters,\cite{boom}
namely, well depth $\epsilon_{\rm LJ} = 10.22$~K and hard core
radius $\sigma_{\rm LJ} = 2.556$~\AA. In order to recover the
correct results for bulk liquid, the screening distance $h_{\rm
OP}$ was adjusted so that the integral of $V^{\rm OP}_l(r)$ over
the whole three-dimensional space be equal to the value of $b_4$
given by Eq.\ (\ref{values})
\begin{eqnarray}
b_4~&&= \int {d{\bf r} V^{\rm OP}_l(r) } \nonumber\\
&&= \frac{32 \pi}{21} \sigma^3_{\rm LJ} \epsilon_{\rm LJ}
\biggr[ \frac{8}{3} \left(\frac{\sigma_{\rm LJ}}{h_{\rm OP}}
\right)^9 - 5 \left(\frac{\sigma_{\rm LJ}}{h_{\rm OP}}\right)^3
\biggr] \;. \label{be}
\end{eqnarray}
This procedure led to $h_{\rm OP} = 2.376\,728$~\AA.

The $\bar{\rho}({\bf r})$ is the ``coarse-grained density'' defined
as the straight average of $\rho({\bf r})$ over a sphere centered 
at ${\bf r}$ and with a radius equal to the screening distance
$h_{\rm OP}$
\begin{equation}
\bar{\rho}({\bf r}) = \int { d{\bf r}\prime \, \rho({\bf r}\prime)
\,\cal{W}(\mid {\bf r} - {\bf r}\prime \mid) } \;, \label{coarse}
\end{equation}
where $\cal{W}(\mid {\bf r} - {\bf r}\prime \mid)$ is taken as the
normalized step function
\begin{eqnarray}
\cal{W}(\mid{\bf r} - {\bf r}\prime \mid)~&&= \frac{3}{4 \pi h_{\rm
OP}^3} \Theta(\,h_{\rm OP}\,- \mid {\bf r} - {\bf r}\prime \mid)
\nonumber\\
&&\: = \left\{
\begin{array}{lll}
\frac{3}{4 \pi h_{\rm OP}^3} & {\rm if} & \mid {\bf r} - {\bf
r}\prime \mid \; \leq h_{\rm OP} \;, \\
& & \\
0 & {\rm if} & \mid {\bf r} - {\bf r}\prime \mid \; > h_{\rm OP}
\;. \\
\end{array} \right. \label{the}
\end{eqnarray}

\subsection{Hartree like equation}
\label{sec:data}

In the case of curved geometries the Eq.\ (\ref{hair0}) takes the
form
\begin{eqnarray}
-\frac{\hbar^2}{2 m} \:&& \left(\,\frac{d^2}{dr^2}
+ \frac{D-1}{r}\,\frac{d}{dr} \right) \,\sqrt{\rho(r)} \nonumber\\
&&+~\biggr[\,V_H(r) + U_{\rm sub}(r) \biggr] \,\sqrt{\rho(r)}
= \mu \, \,\sqrt{\rho(r)} \;, \label{hairs}
\end{eqnarray}
with $D=2$ for cylindrical systems and $D=3$ for spherical ones.
The equation for planar systems is obtained by setting $D=1$ and
assuming that $r$ represents the coordinate perpendicular to the
plane of symmetry (usually denoted $z$). The Hartree potential
derived by applying Eq.\ (\ref{harp}) to the Skyrme-DF reads
\begin{eqnarray}
V^{\rm Sky}_H(r) =&& b_4\,\rho(r) + \frac{\gamma_4+2}{2}\,c_4\,
\rho^{\gamma_4+1}(r) \nonumber\\
&& - 2\,d_4\,\left(\,\frac{d^2}{dr^2} + \frac{D-1}{r}\,\frac{d}{dr}
\right) \,\rho(r) \;. \label{harp0}
\end{eqnarray}
The expression for $V_H(r)$ derived in the OP-NLDF approach for
cylindrical systems is given in the Appendix of Ref.\ \onlinecite{%
less}. The corresponding one for spherical systems is provided in
the Appendix A of the present work. It is worthwhile to notice that
we expressed this quantity in a rather simple compact form similar
to Eq.\ (2.16) written in Ref.\ \onlinecite{ccst} for planar films
instead of adopting the expansion in terms of Legendre polynomials
proposed in Ref.\ \onlinecite{mash}. The latter procedure is more
appropriate for studying excitations of a given system. We shall
mainly report results for curved geometries obtained by setting
$U_{\rm sub}(r) \equiv 0$.

In the case of cylindrical symmetry Eq.\ (\ref{hairs}) is solved
for a fixed number of particles per unit length $L$
\begin{equation}
n_\lambda = N/L = 2\,\pi \int^\infty_0 {r\;dr\;\rho(r)} \;,
\label{number1}
\end{equation}
while for helium spheres the constrain is directly the number of
particles
\begin{equation}
N = 4\,\pi \int^\infty_0 {r^2\;dr\;\rho(r)} \;.
\label{number2}
\end{equation}
Typical density profiles are displayed in Fig. 2 of Ref.
\onlinecite{less} for cylinders and in Fig. 1 of Ref. \onlinecite{%
sat1} for spheres.

We shall first test whether the size of the considered systems of
liquid $^4$He is large enough to expect asymptotic behaviors. Next,
we shall focus our attention on the surface energy and the width of
the interface.

\section{Asymptotic global behavior of the systems}
\label{sec:large}

In the case of spherical systems, it has become customary to define
for each moment $<r^k>$ an equivalent uniform radius $R_k$. This
quantity is equal to the radius of a uniformly occupied sphere (of
density $\rho_u$) with the same momentum of order $k$ given by the
true density distribution. It is possible to extend this idea to
the cylindrical geometry. So, in general, an equivalent uniform
radius $R_k$ associated to a momentum $r^k$ given by
\begin{eqnarray}
< r^k_u >_D \,&&= \frac{2^{D-1}\,\pi\,L^{3-D} \int^{R_k}_0 {dr\;
r^{k+D-1}\;\rho_u}}{2^{D-1}\,\pi\,L^{3-D} \int^{R_k}_0 {dr\;r^{D-1}
\;\rho_u}} \nonumber\\
&&= \frac{ \int^{R_k}_0 {dr\;r^{k+D-1}\;\rho_u}}{\int^{R_k}_0
{dr\;r^{D-1}\;\rho_u}} = \frac{D}{k+D}\,R^k_k \;, \label{mean0}
\end{eqnarray}
is determined by matching the result of this integral with the
$k$-moment of the true radial distribution of the system
\begin{equation}
< r^k(N) >_D \,= \frac{2^{D-1}\,\pi\,L^{3-D}}{N} 
\int^\infty_0 {dr\;r^{k+D-1}\;\rho(r)} \;. \label{mean}
\end{equation}
This condition leads to the generalized Ford-Wills moments (see
Eq.\ (1) in Ref. \onlinecite{wills})
\begin{equation}
R_k = \biggr[ \frac{k+D}{D} < r^k(N) >_D \biggr]^{1/k} \;.
\label{radius0}
\end{equation}

\subsection{Cylindrical systems}
\label{sec:cylinder1}

One expects that very thick cylinders should tend to exhibit
features of bulk liquid at saturation conditions.
In the asymptotic limit of very large cylinders one expects that
Eq.\ (\ref{number1}) will reduce to
\begin{equation}
N/L = 2\,\pi \int^{R_0}_0 {r\;dr\;\rho_0} = \pi\,R^2_0\;\rho_0 \;,
\label{number10}
\end{equation}
where $\rho_0$ is the saturation density of bulk $^4$He quoted in
Table\ \ref{table1} and $R_0$ is the radius of the uniformly filled
cylinder. Since in the limit of very large cylinders the equivalent
radius $R_k$ tends toward $R_0$ defined by Eq.\ (\ref{number10}),
one gets the following asymptotic law
\begin{equation}
\lim_{n_\lambda \to \infty} \biggr\{\frac{[<r^k(n_\lambda)>]^{1/k}}
{n^{1/2}_\lambda} \biggr\} = \sqrt{\frac{1}{\pi \rho_0}}
\left(\frac{2}{k+2}\right)^{1/k} \;. \label{radius3}
\end{equation}
Figure \ref{moment1} shows how the normalized moments $<r>$ and
$<~r^2 >$ attain their asymptotic values as a function of the
expansion parameter $\nu_{\rm cyl} = n^{-1/2}_\lambda$. One may
conclude that the results given by Eq.\ (\ref{radius3}) are safely
reached for $n_\lambda \simeq 70$~\AA$^{-1}$.

\subsection{Spherical systems}
\label{sec:sphere1}

From a similar analysis to that performed for cylinders, one can
state that for very large spheres Eq.\ (\ref{number2}) will become
\begin{equation}
N = 4\,\pi \int^{R_0}_0 {r^2\;dr\;\rho_0} = \frac{4\,\pi}{3}\,
R^3_0\;\rho_0 \;, \label{number20}
\end{equation}
where $R_0$ is now the radius of the uniformly filled sphere.
Furthermore, since in this asymptotic limit $R_k$ tends toward
$R_0$ defined by Eq.\ (\ref{number20}), the following law is
obtained
\begin{equation}
\lim_{N \to \infty} \biggr\{\frac{[<r^k(N)>]^{1/k}}{N^{1/3}}
\biggr\} = \left(\frac{3}{4 \pi \rho_0}\right)^{1/3}
\left(\frac{3}{k+3}\right)^{1/k} \;. \label{radius4}
\end{equation}
Figure \ref{moment2} shows how the normalized moments $< r >$ and
$<~r^2 >$ attain their asymptotic values as a function of $\nu_{\rm
sph} = N^{-1/3}$. The results given by Eq.\ (\ref{radius4}) are
reached at $N \simeq 2000$.

In summary, from values of the normalized moments displayed in
Figs.\ \ref{moment1} and \ref{moment2} it is clear that the largest
systems examined in the present work have, in practice, reached the
asymptotic global behavior.

\section{Analysis of results}
\label{sec:study}

\subsection{Surface tension}
\label{sec:tension}

The evolution of the central density, $\rho_c$, for cylinders with
increasing $n_\lambda$ has been explored in Ref. \onlinecite{less}.
It was found that after a certain value of $n_\lambda$ the central
density becomes bigger than the saturation density of infinite
helium matter, $\rho_c > \rho_0$, giving rise to a squeezing
effect. This phenomenon has been previously found in calculations
carried out for atomic nuclei (see Figs. 1a and 1b in Ref.\
\onlinecite{trend0}) as well as  for spherical clusters of $^4$He
(see Fig. 3 in Ref.\ \onlinecite{sat1}) and it is known as the
``leptodermous'' behavior. It was demonstrated\cite{less} that for
cylinders the squeezing effect, {\it i.e.} the difference $\rho_c -
\rho_0$, vanishes for $n_\lambda \to \infty$ according to the law
\begin{equation}
\epsilon_{\rm cyl} = \frac{\rho_c - \rho_0}{\rho_0}
= \frac{\sigma_\infty}{\cal{K}}\,\sqrt{\frac{\pi}{\rho_0}}\,
n^{-1/2}_\lambda \:, \label{epic1}
\end{equation}
which is similar to
\begin{equation}
\epsilon_{\rm sph} = \frac{\rho_c - \rho_0}{\rho_0}
= 2\,\frac{\sigma_\infty}{\cal{K}}\left(\frac{4\,\pi}{3\,\rho^2_0}
\right)^{1/3}\,N^{-1/3} \:, \label{epic2}
\end{equation}
corresponding to $^4$He spheres (see the asymptotic limit of Eq.\
12 in Ref.\ \onlinecite{sat1}).

In the present work we shall describe the evolution of the surface
tension when the size of the $^4$He systems is increased. From
thermodynamic considerations one gets that
\begin{equation}
dE_{\rm gs} = -P\,dV + \sigma_A\,dA + \mu\,dN \;, \label{en1}
\end{equation}
where $V$ is the volume of the system, $A$ the surface area
enclosing $V$, and $\sigma_A$ is the surface tension. For both
curved geometries the grand free energy takes the form
\begin{equation}
\Omega = E_{\rm gs} - \mu\,N = -P\,V + \sigma_A\,A \;,
\label{omega20}
\end{equation}
while $P=0$ for planar slabs. In the following lines we shall
examine the different cases.

\subsubsection{Cylindrical systems}
\label{sec:cylinder2}

In Ref.\ \onlinecite{less} it has been shown that for cylinders the
relationship between the grand free energy per unit length and the
surface tension becomes
\begin{equation}
\frac{\Omega}{L} = \frac{E_{\rm gs} - \mu\,N}{L} = (\,e - \mu\,)\,
n_\lambda = \pi R_s\,\sigma_A \;. \label{sigh}
\end{equation}
Here $R_s$ is the sharp mean radius defined by
\begin{equation}
N/L = 2\,\pi \int^{R_s}_0 {r\;dr\;\rho_c} = \pi\,R^2_s\;\rho_c \;,
\label{number11}
\end{equation}
and it is related to $R_0$ by
\begin{equation}
R_s = R_0\,\sqrt{\frac{1}{1+\epsilon_{\rm cyl}}} \;. \label{radius}
\end{equation}
Furthermore, it has been demonstrated in Ref.\ \onlinecite{less}
that for large cylindrical systems (in the sense of big $R_s$) the
total energy per particle may be expressed as
\begin{eqnarray}
e = \frac{E_{\rm gs}}{N} = &&e_B + 2\,\sigma_\infty\,
\sqrt{\frac{\pi}{\rho_0}}\,n^{-1/2}_\lambda
+ \left(\,\varepsilon_\ell - \frac{\pi\,\sigma^2_\infty}{2\,
\rho_0\,\cal{K}}\,\right) n^{-1}_\lambda \nonumber\\
&&+ \cdots \;, \label{foot}
\end{eqnarray}
where $\sigma_\infty$ and $\varepsilon_\ell$ are, respectively, the
asymptotic surface tension and the residual energy per unit length
defined in Ref. \onlinecite{less}. A formula for the surface
tension $\sigma_A$ may be derived by taking into account Eq.\ (5.7)
in Ref.\ \onlinecite{less}
\begin{equation}
\pi R_s\,\sigma_A
= -\,n^2_\lambda\,\frac{d e}{d n_\lambda} \;, \label{sigfth}
\end{equation}
and the present relationship (\ref{sigh}). After keeping all terms
up to 2-nd order in $\nu_{\rm cyl}=n^{-1/2}_\lambda$ one gets
\begin{eqnarray}
\sigma_A~&&= \frac{2\,\Omega}{A_{\rm cyl}}
= \frac{2\,(\,e - \mu\,)\,N}{2\,\pi R_s\,L}
= -\,\frac{n^2_\lambda}{\pi\,R_s}\,\frac{de}{dn_\lambda}
\nonumber\\
&&= \sigma_\infty + \sqrt{\frac{\rho_0}{\pi}}\,
\varepsilon_\ell\,n^{-1/2}_\lambda
+ \frac{\sigma_\infty}{2\,\cal{K}}\,\left(\,\varepsilon_\ell - 
\frac{3}{4}\,\frac{\pi\,\sigma^2_\infty}{\rho_0\,\cal{K}}\,\right)
\,n^{-1}_\lambda \;. \nonumber\\
\label{sigf1}
\end{eqnarray}
This expression indicates that the asymptotic result
$\sigma_A(n_\lambda \to \infty) = \sigma_\infty$, which may be
identified with the experimental\cite{gest,isis,raw} values of the
surface tension $\sigma_{\rm exp}$ quoted in Table\ \ref{table1},
should be attained linearly in the parameter $\nu_{\rm cyl} =
n^{-1/2}_\lambda$ with a slope proportional to $\varepsilon_\ell$
\begin{equation}
\sigma_A \simeq \sigma_\infty + \sqrt{\frac{\rho_0}{\pi}}\,
\varepsilon_\ell\, n^{-1/2}_\lambda \;.
\label{sigf2}
\end{equation}
The values of $\varepsilon_\ell$ obtained from the fits of energies
per particle to Eq.\ (\ref{foot}) are listed in Table \ref{table1}.
Since these results are positive one expects an overshoot (i.e., a
region where $\sigma_A(n_\lambda) > \sigma_\infty$) for large
cylinders.

\subsubsection{Spherical systems}
\label{sec:sphere2}

In the case of spherical systems, upon starting from Eqs.\ (\ref{%
en1}) and (\ref{omega20}) one arrives at
\begin{equation}
d\Omega = d(E_{\rm gs} - \mu\,N) = -P\,dV + \sigma_A\,dA - N\,d\mu
\;. \label{sigh21}
\end{equation}
At equilibrium, for a fixed $\mu$, the virtual work by changing the
radius of the sphere should vanish
\begin{equation}
-P\,dV + \sigma_A\,dA = -P\,4\pi\,R^2_s\,dR_s
+ \sigma_A\,8\pi\,R_s\,dR_s = 0 \;. \label{sigh22}
\end{equation}
This equation leads to
\begin{equation}
P = \frac{2\,\sigma_A}{R_s} \;, \label{sigh24}
\end{equation}
where the sharp mean radius is
\begin{equation}
R_s = R_0\,\left(\frac{1}{1+\epsilon_{\rm sph}}\right)^{1/3} \;,
\label{radius2}
\end{equation}
with $R_0$ defined by Eq.\ (\ref{number20}). Then by using
(\ref{omega20}) one gets
\begin{eqnarray}
\Omega~&&= -P\,V + \sigma_A\,A
=-\left(\frac{2\,\sigma_A}{R_s}\right)\left(\frac{4\pi}{3}\right)
\,R^3_s + \sigma_A\,4\pi\,R^2_s \nonumber\\
&&= \frac{4\pi}{3}\,R^2_s\,\sigma_A \;. \label{sigh25}
\end{eqnarray}
The total energy per particle can be expanded in the following way
[cf. Eq.\ (13) in Ref.\ \onlinecite{sat1}]
\begin{eqnarray}
e = &&\frac{E_{\rm gs}}{N} = e_B + \sigma_\infty\,
\left(\frac{36\,\pi}{\rho^2_0}\right)^{1/3}\,N^{-1/3}
\nonumber\\
&&+ \biggr[\,a_c - 2\,\frac{\sigma^2_\infty}{\cal{K}}
\left(\frac{4\,\pi}{3\,\rho^2_0}\right)^{2/3}\,\biggr]\,N^{-2/3}
+ \cdots \;. \label{foots}
\end{eqnarray}
Hence, for spherical systems one gets
\begin{eqnarray}
\sigma_A~&&= \frac{3\,\Omega}{A_{\rm sph}}
= \frac{3\,(\,e - \mu\,)\,N}{4\,\pi R^2_s}
= -\,\frac{3\,N^2}{4\pi\,R^2_s}\,\frac{de}{dN} \nonumber\\
&&\simeq \sigma_\infty + a_c\,\left(\frac{2\,\rho^2_0}{9\,\pi}
\right)^{1/3}\,N^{-1/3} \;. \label{sigh26}
\end{eqnarray}
The values of $a_c$ determined by fitting energies per particle of
drops with $N \geq 300$ to Eq.\ (\ref{foots}) are quoted in Table
\ref{table1}. Since these results are positive, overshoots of
$\sigma_A$ should be expected. For the sake of comparison prior
results reported in Refs.\ \onlinecite{sat1} and \onlinecite{sat2}
obtained by using the Skyrme-DF and considering systems with $N
\leq 728$ are also included in Table \ref{table1}. All these data
are in good agreement.

\subsubsection{Comparison of results for regular geometries}
\label{sec:study1}

Let us now summarize the results for the surface tension for the
three regular geometries. Instead of showing the surface tension
separately for each geometry as a function of the appropriate
expansion parameter $\nu$, we find it more interesting to present
all the data in a unified picture. In order to achieve this goal it
is useful to consider that for all three symmetries (planar,
cylindrical and spherical) the final trend of the energy per
particle toward the asymptotic value, i.e., the bulk $e_B$, is
mainly determined by the linear term in the corresponding expansion
parameter. Therefore, it becomes reasonable to adopt the energy
difference $e - e_B$ as an appropriate common abscissa. In order to
facilitate the forthcoming discussion we recall that for a
symmetric planar slab (see, e.g., Ref.\ \onlinecite{eu}) the energy
per particle may be expressed as an expansion in terms of inverse
coverage $n^{-1}_c=A_{\rm slab}/N$
\begin{equation}
e = \frac{E_{\rm gs}}{N} = e_B + 2\,\sigma_\infty\,n^{-1}_c
+ \gamma_c\,n^{-3}_c + \cdots \;, \label{ene1}
\end{equation}
and the total surface tension becomes
\begin{eqnarray}
\sigma^{(\rm tot)}_A~&&= \frac{\Omega}{A_{\rm slab}}
= \frac{E_{\rm gs} - \mu\,N}{A_{\rm slab}}
= \frac{(\,e - \mu\,)\,N}{L^2} \nonumber\\
&&= -\,n^2_c\,\frac{de}{dn_c} \simeq 2\,\sigma_\infty + 3\,\gamma_c
\,n^{-2}_c \;. \label{sigh0}
\end{eqnarray}

Values for the surface tension obtained from calculations carried
out by using the Skyrme-DF are displayed in Fig.\ \ref{tension1},
while Fig.\ \ref{tension2} shows the behavior of $\sigma_A$
evaluated by using the finite range OP-NLDF. The main features
exhibited by both these drawings are similar. In particular, for
curved geometries the surface tension tends toward $\sigma_\infty$
showing the predicted overshoot, which is similar to that
previously found for $\rho_c$ (see Fig. 6 in Ref. \onlinecite{less}
and Fig. 3 in Ref. \onlinecite{sat1}). Furthermore, in both figures
the final trends toward $\sigma_\infty$ match the linear asymptotic
expressions:

a) for cylindrical systems
\begin{equation}
\sigma_A \simeq \sigma_\infty + \frac{\rho_0\,\varepsilon_\ell}
{2\,\pi\,\sigma_\infty}\,(e - e_B) \;; \label{sigh1}
\end{equation}

b) for spherical systems
\begin{equation}
\sigma_A \simeq \sigma_\infty + \frac{\rho_0\,a_c}{3\,
\pi\,\sigma_\infty}\,\left(\frac{\pi\,\rho_0}{6}\right)^{1/3}
\,(e - e_B) \;. \label{sigh2}
\end{equation}
On the other hand, the data for planar slabs attain the asymptotic
value in a smooth way with zero slope, that is without any linear
term. This behavior may be understood on the basis of the fact that
upon starting from Eq.\ (\ref{sigh0}) one gets

c) for planar systems
\begin{equation}
\sigma_A = \frac{\sigma^{(\rm tot)}_A}{2} \simeq \sigma_\infty
+ \frac{3\,\gamma_c} {8\,\sigma^2_\infty}\,(e - e_B)^2 \;.
\label{sigh01}
\end{equation}
Here $\sigma_A$ is the surface tension at each free interface.
The values obtained with the OP-NLDF approach match this law with
$\gamma_c=-1.44 \times 10^{-3}$~K/\AA$^6$ listed in Table 2 of
Ref. \onlinecite{eu}, while the trend of results provided by the
Skyrme-DF is even flatter suggesting a $(e - e_B)^n$ law with $n
\geq 3$. The latter behavior is due to the fact that in such a
mean-field approximation $\gamma_c \simeq 0$ because the
Lennard-Jones potential is not included in the theory (cf. the
Table 2 and the discussion in Ref. \onlinecite{eu}).

\subsubsection{Relation between $a_c$ and $\varepsilon_\ell$}
\label{sec:relation}

We shall now search for a relation between $a_c$ and
$\varepsilon_\ell$. According to the Droplet Model (DM)\cite{%
trend0,trend1} the total ground-state energy of a large system may
be written as a sum of volume, surface, and curvature terms
\begin{equation}
E_{\rm gs} = E_v + E_s + E_c \;. \label{energy0}
\end{equation}
The surface and curvature energies for cylindrical systems may be
written as (see Eq.\ (4.8) in Ref.\ \onlinecite{less})
\begin{eqnarray}
E_s({\rm cyl})~&&= 2\,\pi\,R_0\,L \int^\infty_0 \{\rho(r){\cal{H}}
[\rho,{\bf \nabla}\rho] - \rho_0\,e_0\} dr \nonumber\\
&&= 2\,\pi\,R_0({\rm cyl})\,L\,\sigma_\infty \;, \label{areal}
\end{eqnarray}
\begin{eqnarray}
E_c({\rm cyl})~&&= 2\,\pi\,L \int^\infty_0 \{\rho(r){\cal{H}}[\rho,
{\bf \nabla}\rho] - \rho_0\,e_0\}(r-R_0)dr \nonumber\\
&&= \varepsilon_\ell\,L \;, \label{curve}
\end{eqnarray}
while for spheres according to Eq. (2.11) in Ref.\ \onlinecite{%
trend1} holds
\begin{eqnarray}
E_s({\rm sph})~&&= 4\,\pi\,R^2_0 \int^\infty_0 \{\rho(r){\cal{H}}
[\rho,{\bf \nabla}\rho] - \rho_0\,e_0\} dr \nonumber\\
&&= 4\,\pi\,R^2_0({\rm sph})\,\sigma_\infty = a_s\,N^{2/3} \;,
\label{areals}
\end{eqnarray}
\begin{eqnarray}
E_c({\rm sph})~&&= 8\,\pi\,R_0 \int^\infty_0 \{\rho(r){\cal{H}}
[\rho,{\bf \nabla}\rho] - \rho_0\,e_0\}(r-R_0)dr \nonumber\\
&&= a_c\,N^{1/3} \;. \label{curves}
\end{eqnarray}
In this approach the integrals in Eqs. (\ref{areal}) and
(\ref{areals}) are associated to $\sigma_\infty$, and it was
verified numerically that yield equal results. The integrals in
Eqs. (\ref{curve}) and (\ref{curves}) may be considered as a higher
order moment of the former integrals, hence, it becomes plausible
to suppose that they should exhibit similar features. Therefore,
one would be able to write
\begin{equation}
E_c({\rm cyl}) = \pi\,L\,\lambda_\infty = \varepsilon_\ell\,L \;,
\label{curves1}
\end{equation}
and
\begin{equation}
E_c({\rm sph}) = 4\,\pi\,R_0({\rm sph})\,\lambda_\infty
= a_c\,N^{1/3} \;. \label{curve1}
\end{equation}
Here $\lambda_\infty$ is twice the asymptotic value of the
integrals in Eqs. (\ref{curve}) and (\ref{curves}) and may be
interpreted as the asymptotic coefficient for the curvature energy.
It is worthy of notice that the curvature energy may be expressed
in terms of the area and curvature of the cylindrical and spherical
surfaces. According to a theorem of Euler the average curvature at
a given point of a surface is defined as\cite{row}
\begin{equation}
{\cal{C}} = \frac{1}{2}\,\left(\frac{1}{R_1}+\frac{1}{R_2}\right)
\;, \label{curl}
\end{equation}
where $R_1$ and $R_2$ are the radii of curvature along any two
orthogonal tangents. Hence, for wide cylinders one gets
\begin{equation}
{\cal{C}}_{\rm cyl} = \frac{1}{2}\,\left(\frac{1}{R_\varphi}
+ \frac{1}{R_z \rightarrow \infty} \right)
= \frac{1}{2\,R_0({\rm cyl})} \;; \label{curl1}
\end{equation}
for large spheres
\begin{equation}
{\cal{C}}_{\rm sph} = \frac{1}{2}\,\left(\frac{1}{R_\varphi} +
\frac{1}{R_\theta}\right) = \frac{1}{R_0({\rm sph})} \;,
\label{curl2}
\end{equation}
while for planar slabs it holds ${\cal{C}}_{\rm slab} = 0$. So,
one may write
\begin{equation}
E_c({\rm cyl}) = 2\,\pi\,R_0({\rm cyl})\,L\,\frac{1}{2\,R_0({\rm
cyl})}\,\lambda_\infty = A_{\rm cyl}\,{\cal{C}_{\rm cyl}}\,
\lambda_\infty \;, \label{curve2}
\end{equation}
and
\begin{equation}
E_c({\rm sph}) = 4\,\pi\,R^2_0({\rm sph})\,\frac{1}{R_0({\rm sph})}
\,\lambda_\infty = A_{\rm sph}\,{\cal{C}_{\rm sph}}\,\lambda_\infty
\;. \label{curves2}
\end{equation}
The relation between $a_c$ and $\varepsilon_\ell$ may be derived
from the ratio of these equations
\begin{equation}
\frac{E_c({\rm sph})}{E_c({\rm cyl})}
= \frac{A_{\rm sph}\,{\cal{C}_{\rm sph}}\,\lambda_\infty}
{A_{\rm cyl}\,{\cal{C}_{\rm cyl}}\,\lambda_\infty}
= \frac{4\,R_0({\rm sph})}{L}
= \frac{a_c\,N^{1/3}}{\varepsilon_\ell\,L} \;, \label{curve01}
\end{equation}
which yields
\begin{equation}
\frac{a_c}{\varepsilon_\ell} = \frac{4\,R_0({\rm sph})}{N^{1/3}}
= 4\,\left(\frac{3}{4\,\pi\,\rho_0}\right)^{1/3} \;,
\label{curve02}
\end{equation}
leading to
\begin{equation}
\left(\frac{\pi\,\rho_0}{48}\right)^{1/3}\,\frac{a_c}
{\varepsilon_\ell} = 1 \;. \label{curve03}
\end{equation}
The values of this relation calculated with parameters determined
from both examined DF are included in Table\ \ref{table1}. For the
Skyrme-DF the result was $0.99$ in excellent agreement with the DM
prediction. The ratio yielded by the OP-NLDF, i.e. $1.10$, differs
slightly from unity. This is indeed to be expected when the DF is
not only written in terms of $\rho(r)$ but has explicit information
on $r$ as happens with Lennard-Jones potential. In such a case
there is an additional curvature contribution which in the approach
adopted here is embedded in the coefficient $a_c$.\cite{trend1} 

Upon starting from Eqs.\ (\ref{sigf2}) and (\ref{sigh26}), in both
cases, one may express the excess of surface energy in terms of the
curvature and the quantity $\lambda_\infty$ getting
\begin{eqnarray}
(\sigma_A - \sigma_\infty)_{\rm cyl}~&&\simeq
\sqrt{\frac{\rho_0}{\pi}}\,\varepsilon_\ell\, n^{-1/2}_\lambda
= \frac{\varepsilon_\ell}{\pi\,R_0({\rm cyl})} \nonumber\\
&&= \frac{2\,\varepsilon_\ell}{\pi}\,{\cal{C}}_{\rm cyl}
= 2\,\lambda_\infty\,{\cal{C}}_{\rm cyl} \;,
\label{sigh10}
\end{eqnarray}
and
\begin{eqnarray}
&&(\sigma_A - \sigma_\infty)_{\rm sph} \simeq a_c\,\left(\frac{2\,
\rho^2_0}{9\,\pi}\right)^{1/3} N^{-1/3} \nonumber\\
&&= \left(\frac{\pi\,\rho_0}{6}\right)^{1/3} \frac{a_c}{\pi\,
R_0({\rm sph})}
= \left(\frac{\pi\,\rho_0}{6}\right)^{1/3}\,\frac{a_c}{\pi}\,
{\cal{C}}_{\rm sph} \nonumber\\
&&= 2\,\lambda_\infty\,{\cal{C}}_{\rm sph} \;. \label{sigh20}
\end{eqnarray}
By taking into account these results one arrives at
\begin{eqnarray}
\frac{(\sigma_A - \sigma_\infty)_{\rm sph}\,A_{\rm sph}}
{(\sigma_A - \sigma_\infty)_{\rm cyl}\,A_{\rm cyl}} =\frac{A_{\rm
sph}\,{\cal{C}}_{\rm sph}}{A_{\rm cyl}\,{\cal{C}}_{\rm cyl}}
= \frac{E_c({\rm sph})}{E_c({\rm cyl})} \;. \label{sigh31}
\end{eqnarray}
This means that the ratio of the excess of surface energies is
equal to the ratio of curvature energies.

\subsection{Thickness at the free interface}
\label{sec:widths}

All the calculated results for the surface thickness $W$ of free
$^4$He systems are plotted together in Fig.\ \ref{width} as a
function of $e - e_B$. Let us first look at values obtained by
applying the Skyrme-DF. In this case we must state a word of
caution because our results for spheres differ noticeably from
those listed in the fifth column of Table I in Ref.\
\onlinecite{sat1}. We attribute this difference to a misprint in
Table I of Ref.\ \onlinecite{sat1} in view of the fact that the
tabulated values are also inconsistent with the thickness inferred
from the profiles plotted in Fig. 1 of the same paper. This point
is stressed because due to that mistake the range of $W$ ascribed
to calculations with the Skyrme-DF for spheres with $N \leq 728$
quoted in the summary of Table III in Ref.\ \onlinecite{dali} is
incorrect, actually it is $6.6 \lesssim W \lesssim 7.4$~\AA.

Turning to the present results, the displayed data indicate that
the asymptotic surface width obtained with the zero-range
functional does not depend on the geometry of the helium system. In
all the cases it attains $W_\infty \simeq 7$~\AA, in agreement with
the asymptotic result for an infinite system reported in Ref.\
\onlinecite{sat1}.

The results obtained from calculations carried out with the OP-NLDF
are smaller than those yielded by the Skyrme-DF, but exhibit a
similar trend as a function of $e-e_B$ and the asymptotic width is
also independent of the geometry being $W_\infty \simeq 5.9$~\AA.
For large systems the OP-NLDF thickness is close to that reported
for droplets in Table\ II of Dalfovo et al.\cite{dali} These
authors have performed calculations for large systems ($1000 \leq N
\leq 10000$, i.e., $-6.4 \lesssim e \lesssim -5.4$ K) with the more
elaborated OT-NLDF getting $W \simeq 5.7$~\AA, as may be seen in
the drawing.

The values obtained from Monte Carlo simulations for free planar
slabs\cite{gall,vale} and spherical drops\cite{gall} are also
plotted in Fig.\ \ref{width}. However, in particular, the data
evaluated by using glue-SWF\cite{gall} exhibit a too large spread
to make possible any meaningful systematics.

The comparison between the theoretical results mentioned above and
the experimental data is performed in the same Fig.\ \ref{width}.
The data obtained in the measurement of large droplets published in
Ref.\ \onlinecite{dali} are represented by the reported mean
thickness $W = 6.4 \pm 1.3$~\AA. In the case of Ref.\ \onlinecite{%
penal} only the two data included by the authors in the abstract
are shown. Since these values correspond to rather broad planar
films, we plotted them schematically close to origin for the
abscissa. A similar criterion was adopted for the value of Ref.\
\onlinecite{sill}. As can be seen in this figure, all the results
provided by finite range functionals for the largest systems are
consistent with the most recent and precise experimental data.
\cite{penal}

The measurements of Refs.\ \onlinecite{sill} and \onlinecite{penal}
have been performed for planar films of $^4$He adsorbed onto a
substrate of Si, while the calculations mentioned above have been
carried out for free planar slabs. Therefore we completed the
investigation exploring how much theoretical results change when
helium is adsorbed on a solid surface. In doing so, we include in
this analysis the results of our OP-NLDF estimations of the
thickness at the free interface for $^4$He films adsorbed on
surfaces of Na and Li. These substrates provide the strongest
adsorption potentials among alkali metals.\cite{chics,ruby} For
example, we show in Fig.\ \ref{width} the values calculated for
stable films with coverage $n_c = 0.34$~\AA$^{-2}$ in the case of
both substrates, i.e., $W = 5.57$ (Na) and $5.45$~\AA~(Li).
\cite{abscissa} These results fit perfectly into the general
pattern exhibited by evaluations performed with NLDF and recent
experimental data.\cite{penal}

\subsubsection{Simple models for the surface thickness}
\label{sec:model}

Let us now show that very simple models provide reasonable
estimations of the interface thickness and its relation with the
surface tension for large systems. For this purpose, it is useful
to start first from the Eq.\ (2.6) of Ref.\ \onlinecite{eu} for
$\sigma_A$ of planar slabs obtained within the Skyrme-DF. It reads
\begin{equation}
\sigma_A = \frac{1}{2} \int^\infty_{-\infty} { dz \biggr[
\frac{\hbar^2}{m} \left(\,\frac{d\sqrt{\rho(z)}}{dz}\right)^2
+ 2\,d_4\,\left(\frac{d\rho(z)}{dz} \right)^2 \biggr] } \;.
\label{sigma0}
\end{equation}
We shall evaluate this integral adopting a crude approximation for
the density profile of a wide slab. Let us consider that from $z=0$
to $z=z_0$ the density is constant and equal to its value at the
center $\rho_c$. Subsequently, for $z>z_0$ begins the falloff.
Assuming that the first stretch of the falloff up to $z=z_\ell$ is
linear, provided that $\rho(z_\ell) < \rho_c/10$, one may write
\begin{equation}
\rho(z)= \rho_c\,\left(\,1 - \frac{z-z_0}{\tilde{W}}\,\right)
= \rho_c\,\left(\,1 - \frac{4\,(z-z_0)}{5\,W}\,\right) \;.
\label{rho00}
\end{equation}
Next, for $z \geq z_\ell$, the falloff of the density profile
follows the asymptotic exponential law (see, for instance, Eqs.\
(7) and (9) in Ref.\ \onlinecite{edward})
\begin{equation}
\rho(z) = \frac{\rho_c}{\alpha}\,\exp{[-2\,B\,(z-z_\ell)]} \;,
\label{rho01}
\end{equation}
with $B=\sqrt{-2m\,e_B/\hbar^2} = 1.09$~\AA$^{-1}$. The parameters
$z_\ell$ and $\alpha$ are determined by matching at $z=z_\ell$ the
expressions (\ref{rho00}) and (\ref{rho01}) for $\rho(z)$ as well
as its derivatives $d\rho(z)/dz$. This task is performed in
Appendix B and the result is
\begin{equation}
\sigma_A \simeq \frac{\rho_c}{5\,W} \left(\,\frac{\hbar^2}{m}
+ \,8\,d_4\,\rho_c\,\right) \;. \label{sigma3}
\end{equation}

So, in this crude approach the asymptotic width becomes
\begin{equation}
W_\infty \simeq \frac{\rho_0}{5\,\sigma_\infty} \left(
\,\frac{\hbar^2}{m} + \,8\,d_4\,\rho_0\,\right) = 6.83 \,{\rm
\AA} \;, \label{we}
\end{equation}
lying close to the limit suggested by data displayed in Fig.\
\ref{width}.

A slightly different version of the simple model described above
may provide separate formulas for the surface thickness and tension
in terms of parameters of the Skyrme-DF. As we shall see below,
such expressions are valid for all the analyzed geometries. In this
model, the proposed $\rho (r)$ is the simplest that take into
account  three of the most relevant characteristics of density
profile, the sharp radius of the density distribution $R_s$, the
thickness of the interface $W$, and the ``leptodermous'' behavior
$\rho_c > \rho_0$. We assume that from $r=0$ to $r=R_1=R_s - \beta
\tilde{W}$ the density is constant and equal to $\rho_c=\rho(r=0)$,
we recall that $\tilde{W}=5W/4$. At $r=R_1$ begins a linear falloff
and for $r \geq R_2=R_s + (1 - \beta) \tilde{W}$ the density
vanishes. The parameter $\beta$ depends on the geometry through the
normalization condition given by Eq.\ (\ref{np}), where $N$ should
be suitably written for non spherical systems in terms of $n_{c}$
or $n_{\lambda}$. The obtained formulas for the normalization
condition and the parameter $\beta$ are given in Appendix C.
Subsequently. the ground state energy provided by the Skyrme-DF for
slabs, cylinders, and spheres was expressed by taking into account
the proposed $\rho(r)$. In order to avoid an unphysical divergence
in the integration of the kinetic term caused by the kink of the
adopted density profile we assumed
\begin{equation}
\frac{d\sqrt{\rho(z)}}{dz} \simeq - \frac{4\,\sqrt{\rho(z)}}{5\,W}
\;. \label{kine}
\end{equation}
As an example, the energy for cylinders is given in Appendix C.

Upon minimizing the grand free energy (e.g., for cylinders see Eq.\
(\ref{ecyl})) with respect to $W$ and $\epsilon$, and keeping only
the dominant terms, the following relation valid for all three
considered geometries is obtained
\begin{equation}
W\,\sigma_A = \frac{4\,\rho_c}{5} \left(\,\frac{\hbar^2}{m} + \,2
\,d_4\,\rho_c\,\right) \;. \label{sigma4}
\end{equation}
Here, the dominant contribution which is given by the term
proportional to $d_4$ is equal to that previously obtained in Eq.\
(\ref{sigma3}). This procedure also yields an expression for the
asymptotic width in terms of parameters of the functional
\begin{equation}
W_{\infty}=\frac{4}{5}\frac{\sqrt{\left(\hbar^2/m\right)
+2\,d_{4}\rho_{0}}}{\sqrt{-e_B + \frac{b_{4}}{3}\rho_{0}
+\frac{c_4}{\gamma_4+3}\rho^{\gamma_4 +1}_0}} = 6.90\,{\rm \AA} \;.
\end{equation}
This value is in excellent agreement with the asymptotic result
obtained from the complete numerical solutions in the Skyrme-DF
approach. The corresponding result for the asymptotic surface
tension
\begin{equation}
\sigma_{\infty}= \frac{4}{5} \frac{\rho_0}{W_\infty}\,
\left(\frac{\hbar^2}{m} + 2\,d_4\rho_0 \right) = 
0.294\,{\rm K/\AA}^2 \;, \label{sigma5}
\end{equation}
indicates a good degree of self-consistency of this approximation.

In conclusion, we can state that the crude simple models used here
to estimate the surface thickness are able to catch a substantial
part of the physics involved.

\subsubsection{The quantity $W \sigma_A$}
\label{sec:quantity}

By looking at the relation between $W$ and $\sigma_A$ given by
Eqs.\ (\ref{sigma3}) and (\ref{sigma4}), we found it worthy to
examine the product $W\,\sigma_A$ as a function of $e-e_B$. Such a
behavior is shown in Fig.\ \ref{prod}, where for curved geometries
one may observe overshoots similar to that exhibited by the central
density and the surface tension. Since according to Eqs.\
(\ref{sigma3}) and (\ref{sigma4}) the main contribution to $W\,
\sigma_A$ is given by the term containing $d_4$, in this case the
linear departure would be governed by
\begin{equation}
\rho^2_c = \rho^2_0\,(1 + \epsilon)^2 \simeq \rho^2_0\,(1 + 2\,
\epsilon) \;, \label{rho1}
\end{equation}
which leads to
\begin{equation}
W\,\sigma_A \simeq (W\,\sigma_A)_\infty\,(1 + 2\,\epsilon) \;.
\label{prod0}
\end{equation}
This relation yields linear asymptotic expressions, where the slope
is determined by the compressibility $\cal{K}$ which is an input
for the adopted DF approaches:

a) for cylindrical systems
\begin{equation}
W\,\sigma_A \simeq (W\,\sigma)_\infty \biggr[ 1
+ \frac{1}{\cal{K}}\,(e - e_B) \biggr] \;; \label{prod1}
\end{equation}

b) for spherical systems
\begin{equation}
W\,\sigma_A \simeq (W\,\sigma)_\infty \biggr[ 1
+ \frac{4}{3\,\cal{K}}\,(e - e_B) \biggr] \;. \label{prod2}
\end{equation}
These approximations are plotted as dashed lines in Fig.\
\ref{prod}. From this drawing one can realize that these very
simple forms account fairly well for the obtained departures. In
the case of planar systems there is no overshoot.

By combining the expansions for the product $W\,\sigma_A$ and for
the surface tension $\sigma_A$ it is possible to estimate the
departure of $W$ from its asymptotic value:

a) for cylindrical systems
\begin{equation}
W = \frac{(W\,\sigma_A)}{\sigma_A} \simeq W_\infty \biggr[ 1
+ \left(\frac{1}{\cal{K}}-\frac{\rho_0\,\varepsilon_\ell}{2\,\pi\,
\sigma^2_\infty} \right)\,(e - e_B) \biggr] \;; \label{we1}
\end{equation}

b) for spherical systems
\begin{eqnarray}
W~&&= \frac{(W\,\sigma_A)}{\sigma_A} \nonumber\\
&&\simeq W_\infty \biggr\{ 1 + \biggr[ \frac{4}{3\,\cal{K}}
- \frac{\rho_0\,a_c}{3\,\pi\,\sigma^2_\infty}\,\left(\frac{\pi\,
\rho_0}{6}\right)^{1/3} \biggr]\,(e - e_B) \biggr\} \;. \nonumber\\
\label{we2}
\end{eqnarray}
For both geometries there is an important cancellation within the
coefficient of the linear term in these expressions. This fact
explains why for large systems the thickness as a function of $e -
e_B$ is flat in Fig.\ \ref{width}.

\section{Summary}
\label{sec:conclude}

The curvature effects on the thickness of the free interface $W$
and the corresponding surface tension $\sigma_A$ are studied. It is
known that the free spherical systems are those with lowest surface
area to volume ratio which leads to lowest energy per particle, but
nevertheless, we found of interest to examine also other geometries
to have a more complete view of surface properties. In addition,
some general properties of free systems could be extended to
adsorbed ones.

First of all, we tested whether the largest systems have already
reached the critical size for exhibiting a global asymptotic
behavior. From Figs.\ \ref{moment1} and \ref{moment2} one may
realize that the moments $<r^k>$, with $k=1$ and $2$, have attained
the asymptotic values.

In this paper, the results for $\sigma_A$ and $W$ corresponding to
different geometries are presented in a unified scale as a function
of $e-e_B$ allowing a direct comparison. As far as we know, this
procedure has not been previously used.

It was found that for cylindrical and spherical systems the surface
tension presents an overshoot similar to that previously observed
for the central density.\cite{sat1,less} This fact indicates that
the squeezing effect known as the ``leptodermous'' behavior is also
manifested in the surface tension. On the contrary, this feature is
not present in the case of symmetric planar slabs. It is shown that
the behavior of large systems may be well interpreted within the
DM\cite{trend0,trend1} adapted to each geometry. In particular, a
relation between coefficients of the curvature energy is derived.

An analysis of the surface thickness $W$ indicates that for large
systems, i.e., when $e-e_B \lesssim 1$~K, the asymptotic result is
independent of the geometry of the $^4$He system. The departure
from that value is different for each geometry. These features are
common to both examined density functionals.

We should mention that according to a discussion given in the
text, the range $8.8 \lesssim W \lesssim 9.2$~\AA~quoted in Table
III of Ref.\ \onlinecite{dali} as corresponding to droplets (with
$N \leq 728$) studied by utilizing the Skyrme-DF, which was
determined with data taken from Table I of Ref.\ \onlinecite{sat1},
must be replaced by $6.6 \lesssim W \lesssim 7.4$~\AA.

The results obtained with the OP-NLDF for large systems are close
to those calculated with the OT-NLDF for spheres by Dalfovo et al.
\cite{dali} It is shown in Fig.\ \ref{width} that the most recent
experimental data\cite{penal} favor the results yielded by the
finite range density functionals. In addition, the thickness of the
free surface of helium films adsorbed on planar substrates of Na
and Li evaluated with the OP-NLDF matches very well with the
pattern depicted in Fig.\ \ref{width}.

We have also analyzed the product $W \sigma_A$. As expected, for
curved geometries this quantity exhibit an overshoot. However, it
is interesting to note that in this case the departure from the
asymptotic values is mainly determined by the inverse of the
compressibility $\cal{K}$ of bulk $^4$He. Furthermore, this
analysis gives a hint for understanding the flat behavior of $W$ at
small $e-e_B$.

\acknowledgements

This work was supported in part by the Ministry of Culture and
Education of Argentina through Grants PICT-2000-03-08450 from
ANPCYT and No. X103 from University of Buenos Aires.

\appendix
\section*{A. OP-NLDF Hartree mean-field potential}

In this Appendix we compile the relevant expressions derived in the
OP-NLDF approach for a spherical geometry. The explicit form of the
Hartree mean-field potential $V_H(r)$ derived according to Eq.\
(\ref{harp}) becomes
\begin{eqnarray}
V^{\rm OP}_H&&(r) = \frac{\delta E_{sc}[\rho]}{\delta \rho(r)}
= \int d{\bf r}\prime \, \rho(r\prime) \, V^{\rm OP}_l(\mid{\bf
r} - {\bf r}\prime \mid)) \nonumber\\
&&+ \,\frac{c_4}{2}\,[\,\bar{\rho}(r)\,]^{\gamma_4+1} \nonumber\\
&&+ \frac{c_4}{2}(\gamma_4+1) \int d{\bf r}\prime
\rho(r\prime) [\,\bar{\rho}(r\prime)\,]^{\gamma_4}
\,\cal{W}(\mid {\bf r} - {\bf r}\prime \mid) \;. \label{harp1}
\end{eqnarray}
Let us first provide the expressions of the contributions
involving the ``coarse-grained density'' $\bar{\rho}(r)$, {\it
i.e.}, 
\begin{equation}
\bar\rho(r) = \int d{\bf r}\prime \, \rho(r\prime) \,
\cal{W}(\,\mid{\bf r} - {\bf r}\prime \mid) \;, \label{rbar}
\end{equation}
and
\begin{equation}
\bar\rho_V(r) = \int d{\bf r}\prime \, \rho(r\prime) \,
[\,\bar{\rho}(r\prime)\,]^{\gamma_4} \, \cal{W}(\,\mid{\bf r} -
{\bf r}\prime \mid) \;. \label{Vrbar}
\end{equation}
Both these integrals may be cast into the  form
\begin{equation}
\bar{\cal{R}}(r) = \frac{3}{4 \pi h^3_{\rm OP}} \int d{\bf r}\prime
\, {\cal{R}}(r\prime) \, \Theta(\,h_{\rm OP}\,- \mid{\bf r} - {\bf
r}\prime \mid) \;. \label{rrbar}
\end{equation}

After introducing spherical coordinates and taking into account
that the step function is symmetric in the azimuthal angle
$\varphi$, the integration over this variable yields
\begin{eqnarray}
\bar{\cal{R}}(r)~&&= \frac{3}{2\,h^3_{\rm OP}}
\int^{r\prime_{\rm max}}_{r\prime_{\rm min}} r\prime^2\,dr\prime \,
{\cal{R}}(r\prime)\,\int^{\theta_{\rm max}}_0 \sin\theta \, d\theta
\nonumber\\
&&~~~\times \Theta[\,h^2_{\rm OP} - r^2 - r\prime^2
+ 2\,r\,r\prime\,\cos\theta\,] \;. \label{rbar0}
\end{eqnarray}
For the point located at $r=0$ one gets
\begin{eqnarray}
\bar{\cal{R}}(r=0)~&&= \frac{3}{2\,h^3_{\rm OP}}
\int^{r\prime_{\rm max}}_{r\prime_{\rm min}} r\prime^2\,dr\prime\,
{\cal{R}}(r\prime)\,\Theta[\,h^2_{\rm OP} - r\prime^2\,]
\nonumber\\
&&~~~\times \int^\pi_0 \sin\theta \, d\theta \nonumber\\
&& = \frac{3}{h^3_{\rm OP}} \int^{h_{\rm OP}}_0
r\prime^2 \, dr\prime \, {\cal{R}}(r\prime) \;. \label{rbar1}
\end{eqnarray}
For $r>0$ two different cases should be considered: (i) $0 < r <
h_{\rm OP}$ and (ii) $r \geq h_{\rm OP}$. For $0 < r < h_{\rm OP}$
the integral over $r\prime$ should be split into two parts
\begin{eqnarray}
\bar{\cal{R}}&&(r)= \frac{3}{2\,h^3_{\rm OP}} \biggr\{
\int^{h_{\rm OP}-r}_0 + \int^{r+h_{\rm OP}}_{h_{\rm OP}-r} \biggr\}
r\prime^2\,dr\prime\,{\cal{R}}(r\prime) \nonumber\\
&&\times \int^{\theta_{\rm max}}_0 \sin\theta\,d\theta\,
\Theta[\,h^2_{\rm OP} - r^2 - r\prime^2 + 2 r r\prime\,\cos\theta
\,] \;. \label{rbar2}
\end{eqnarray}
Since for the first integral over $r\prime$ the upper angular limit
is $\theta_{\rm max}=\pi$, while for the second integral
$\theta_{\rm max}$ is determined by the condition
\begin{equation}
\cos\theta_{\rm max} = \frac{r^2 + r\prime^2 - h^2_{\rm OP}}
{2\,r\,r\prime} \;, \label{cos1}
\end{equation}
then
\begin{eqnarray}
\bar{\cal{R}}(r)~&&= \frac{3}{2\,h^3_{\rm OP}}
\int^{h_{\rm OP}-r}_0 r\prime^2\,dr\prime \, {\cal{R}}(r\prime)
\int^\pi_0  \sin\theta\,d\theta \nonumber\\
&&+ \frac{3}{2 h^3_{\rm OP}} \int^{r+h_{\rm OP}}_{h_{\rm OP}-r}
r\prime^2\,dr\prime\,{\cal{R}}(r\prime)\,\int^{\theta_{\rm max}}_0
\sin\theta\,d\theta \;. \label{rbar21}
\end{eqnarray}
The integration over $\theta$ leads to a very simple expression
\begin{eqnarray}
\bar{\cal{R}}(r)~&&=\frac{3}{h^3_{\rm OP}} \int^{h_{\rm OP}-r}_0
r\prime^2\,dr\prime\,{\cal{R}}(r\prime) \nonumber\\
&&+ \frac{3}{4\,r\,h_{\rm OP}} \int^{r+h_{\rm OP}}_{h_{\rm OP}-r}
r\prime\,dr\prime\,{\cal{R}}(r\prime)\,\biggr[\,1
- \left(\frac{r-r\prime}{h_{\rm OP}}\right)^2 \, \biggr] \;.
\nonumber\\
\label{rbar22}
\end{eqnarray}
For $r \geq h_{\rm OP}$, there is only one contribution similar to
the second term in Eq.\ (\ref{rbar21})
\begin{eqnarray}
\bar{\cal{R}}(r)~&&= \frac{3}{2\,h^3_{\rm OP}}
\int^{r+h_{\rm OP}}_{r-h_{\rm OP}} r\prime^2\,dr\prime\,{\cal{R}}
(r\prime)\,\int^{\theta_{\rm max}}_0 \sin\theta\,d\theta
\nonumber\\
&&= \frac{3}{4\,r\,h_{\rm OP}} \int^{r+h_{\rm OP}}_{r-h_{\rm OP}}
r\prime\,dr\prime\,{\cal{R}}(r\prime)\,\biggr[\,1
- \left(\frac{r-r\prime}{h_{\rm OP}}\right)^2 \, \biggr] \;.
\nonumber\\ 
\label{rbar3}
\end{eqnarray}

Let us now focus on the integration of the screened LJ potential
contributing to Eq.\ (\ref{harp1})
\begin{eqnarray}
V^{\rm LJScr}_H&&(r) = \int d{\bf r}\prime \, \rho(r\prime) \,
V^{\rm OP}_l(\mid{\bf r} - {\bf r}\prime \mid)) \nonumber\\
&&= V^{\rm OP}_l(h_{\rm OP}) \int d{\bf r}\prime \rho(r\prime)
\left(\frac{R}{h_{\rm OP}}\right)^4 \Theta(h_{\rm OP} - R)
\nonumber\\
&&~~+ \,4 \epsilon \int_{R \ge h_{\rm OP}} d{\bf r}\prime
\rho(r\prime) \biggr[\left(\frac{\sigma}{R}\right)^{12} - \left(
\frac{\sigma}{R}\right)^6\biggr] \;, \label{VHar0}
\end{eqnarray}
with
\begin{equation}
R = \mid{\bf r} - {\bf r}\prime \mid \;. \label{big}
\end{equation}
Here $\sigma$ and $\epsilon$ stand for $\sigma_{\rm LJ}$ and
$\epsilon_{\rm LJ}$, respectively.
In this case one can follow the same procedure as that utilized to
calculate the ``coarse-grained density'' terms. As before there are
three different domains of $r$ to be considered. At $r=0$ the
Hartree potential reads
\begin{eqnarray}
V&&^{\rm LJScr}_H(r=0) = \frac{4\,\pi}{h^4_{\rm OP}}\,
V^{\rm OP}_l(h_{\rm OP}) \int^{h_{\rm OP}}_0 r\prime^6\,
dr\prime\,\rho(r\prime) \nonumber\\
&&+ \,16 \pi \epsilon \int^\infty_{h_{\rm OP}} r\prime^2\,
dr\prime\,\rho(r\prime)\,
\biggr[\left(\frac{\sigma}{r\prime}\right)^{12}
-\left(\frac{\sigma}{r\prime}\right)^6\biggr] \;. \label{VHar1}
\end{eqnarray}
For $0 < r < h_{\rm OP}$ the integral over $r\prime$ becomes
\begin{eqnarray}
V&&^{\rm LJScr}_H(r) = \frac{\pi\,h^2_{\rm OP}}{3\,r}\,
V^{\rm OP}_l(h_{\rm OP}) \nonumber\\
&&~~~~\times \biggr\{ \int^{h_{\rm OP}-r}_0 r\prime\,dr\prime\,
\rho(r\prime)\,\biggr[ \left(\frac{r+r\prime}{h_{\rm OP}}\right)^6
- \left(\frac{r-r\prime}{h_{\rm OP}}\right)^6 \biggr]
\nonumber\\
&&~~~~~~~~+ \int^{r+h_{\rm OP}}_{h_{\rm OP}-r} r\prime\,
dr\prime\,\rho(r\prime)\,\biggr[\,1 - \left(\frac{r-r\prime}
{h_{\rm OP}}\right)^6 \,\biggr] \biggr\} \nonumber\\
&&+ \,\frac{4 \pi \epsilon\,\sigma^2}{r}
\,\int^{r+h_{\rm OP}}_{h_{\rm OP}-r} r\prime\,dr\prime\,
\rho(r\prime)\,\biggr\{\frac{1}{5} \biggr[
\left(\frac{\sigma}{h_{\rm OP}}\right)^{10} \nonumber\\
&&~- \left(\frac{\sigma}{r+r\prime}
\right)^{10} \biggr] - \frac{1}{2} \biggr[ \left(\frac{\sigma}
{h_{\rm OP}}\right)^4 - \left(\frac{\sigma}{r+r\prime}
\right)^4 \biggr]\biggr\} \nonumber\\
&&+ \,\frac{4 \pi \epsilon\,\sigma^2}{r}
\,\int^\infty_{r+h_{\rm OP}} r\prime\,dr\prime\,\rho(r\prime)\,
\biggr\{ \frac{1}{5}
\biggr[ \left(\frac{\sigma}{r-r\prime}\right)^{10} \nonumber\\
&&~ - \left(\frac{\sigma}{r+r\prime}\right)^{10} \biggr]
- \frac{1}{2} \biggr[ \left(\frac{\sigma}
{r-r\prime}\right)^4 - \left(\frac{\sigma}{r+r\prime}
\right)^4 \biggr] \biggr\} \;. \nonumber\\
\label{VHar2}
\end{eqnarray}
Finally, for $r \geq h_{\rm OP}$ one gets
\begin{eqnarray}
V&&^{\rm LJScr}_H(r) = \frac{\pi\,h^2_{\rm OP}}{3\,r}\,
V^{\rm OP}_l(h_{\rm OP}) \nonumber\\
&&~~~\times \int^{r+h_{\rm OP}}_{r-h_{\rm OP}} \rho(r\prime)\,
\biggr[\,1 - \left(\frac{r-r\prime}{h_{\rm OP}}\right)^6 \,\biggr]
\,r\prime\,dr\prime \nonumber\\
&&+ \,\frac{4 \pi \epsilon\,\sigma^2}{r} \biggr\{
\int^{r-h_{\rm OP}}_0 + \int^\infty_{r+h_{\rm OP}} \biggr\}
\,r\prime\,dr\prime \, \rho(r\prime) \nonumber\\
&&~~~ \times \biggr\{ \frac{1}{5}
\biggr[ \left(\frac{\sigma}{r-r\prime} \right)^{10}
- \left(\frac{\sigma}{r+r\prime}\right)^{10} \biggr] \nonumber\\
&&~~~  - \frac{1}{2} \biggr[ \left(\frac{\sigma}
{r-r\prime}\right)^4 - \left(\frac{\sigma}{r+r\prime}
\right)^4 \biggr] \biggr\}
\nonumber\\
&&+ \,\frac{4 \pi \epsilon\,\sigma^2}{r}
\,\int^{r+h_{\rm OP}}_{r-h_{\rm OP}} r\prime\,dr\prime\,
\rho(r\prime)\,\biggr\{\frac{1}{5} \biggr[ \left(\frac{\sigma}
{h_{\rm OP}}\right)^{10} \nonumber\\
&&~- \left(\frac{\sigma}{r+r\prime}
\right)^{10} \biggr] - \frac{1}{2} \biggr[ \left(\frac{\sigma}
{h_{\rm OP}}\right)^4 - \left(\frac{\sigma}{r+r\prime}
\right)^4 \biggr]\biggr\} \;. \nonumber\\
\label{VHar3}
\end{eqnarray}

\section*{B. Simple model I}

Starting from Eqs.\ (\ref{rho00}) and (\ref{rho01}), after imposing
continuity of functions and derivatives one gets
\begin{equation}
\rho(z_\ell) = \rho_c\,\left(\,1 - \frac{4\,(z_\ell-z_0)}{5\,W}\,
\right) = \frac{\rho_c}{\alpha} \;, \label{rho02}
\end{equation}
and
\begin{equation}
\left(\frac{d\rho(z)}{dz}\right)_{z=z_\ell}
= -\,\frac{4\,\rho_c}{5\,W} = -\,\frac{2\,B\,\rho_c}{\alpha} \;.
\label{rho03}
\end{equation}
These conditions lead to
\begin{equation}
\alpha = \frac{5}{2}\,B\,W \;, \label{rho04}
\end{equation}
and
\begin{equation}
z_\ell-z_0 = \frac{5}{4}\,W - \frac{1}{2\,B} = \frac{9}{8}\,W
+ \left( \frac{W}{8} - \frac{1}{2\,B} \right) \;.
\label{rho05}
\end{equation}
Since for large slabs one expects $W > 4/B \simeq 3.7$~\AA, then
$z_\ell- z_0$ would be larger than $9\,W/8$ ensuring $\rho(z_\ell)
< \rho_c/10$, which in turn would support the consistency of the
adopted law for the falloff and the definition of the thickness.

For the evaluation of the derivative one may use the relation
\begin{equation}
\frac{d\sqrt{\rho(z)}}{dz} = \frac{1}{2\,\sqrt{\rho(z)}}\,
\frac{d\rho(z)}{dz} \;. \label{rho2}
\end{equation}
Now, the integral of Eq.\ (\ref{sigma0}) may be split into two
parts
\begin{equation}
\sigma_A = \biggr\{ \int^{z_\ell}_{z_0} + \int^\infty_{z_\ell}
\biggr\} dz \biggr[\,\frac{\hbar^2}{4\,m}\,\frac{1}{\rho(z)} + 2\,
d_4\,\biggr] \left(\,\frac{d\rho(z)}{dz}\right)^2 \;,
\label{sigma1}
\end{equation}
which leads to
\begin{eqnarray}
\sigma_A = \frac{\rho_c}{5\,W} \biggr\{&&
\,\frac{\hbar^2}{m}\,\biggr[\,1 + \ln\left(\frac{5}{2}\,B\,W\right)
\,\biggr] \nonumber\\
&&~+ 8\,d_4\,\rho_c\,\biggr[\,1 - \frac{1}{5\,B\,W}\,\biggr]
\biggr\} \;. \label{sigma2}
\end{eqnarray}
It is possible to verify that for the expected values of $W$ there
is an important cancellation between terms carrying the product
$B\,W$. Hence, Eq.\ (\ref{sigma2}) may be reduced to
\begin{equation}
\sigma_A \simeq \frac{\rho_c}{5\,W} \left(\,\frac{\hbar^2}{m}
+ \,8\,d_4\,\rho_c\,\right) \;. \label{sigma3b}
\end{equation}

\section*{C. Simple Model II}

The evaluation of the normalization condition yields 
\begin{equation}
N=\Omega_D\,\rho_c\,\frac{[R_s+(1-\beta)W ]^{D+1} - [R_s -\beta W
]^{D+1} }{D(D+1)\,L^{D-3}\,W} \;, \label{cn}
\end{equation}
with $\Omega_{D=1}=2$, $\Omega_{D=2}=2\,\pi$, and $\Omega_{D=3}=4\,
\pi$. In the case $D=1$ the "sharp radius" $R_s$ stands for $z_s$.
The parameter $\beta$ becomes:

a) for the planar geometry
\begin{equation}
\beta= \frac{1}{2} \;; \label{cn1}
\end{equation}

b)for cylinders
\begin{equation}
\beta= \frac{1}{2}+\frac{R_s}{W}-\sqrt{\left(\frac{R_s}{W}
\right)^{2}- \frac{1}{12}} \;; \label{cn2}
\end{equation}

c) for spheres
\begin{eqnarray}
\beta&&~= \frac{1}{2}+\frac{R_s}{W}-\frac{1}{2^{1/3}}\biggr[
\left(\frac{R_s}{W}\right)^{3}+\sqrt{\left(\frac{R_s}{W}\right)^{6}
+\frac{1}{2*6^3}}\,\biggr]^{1/3} \nonumber\\
&& + \frac{2^{1/3}}{12} \biggr[\left(\frac{R_s}{W}\right)^{3}
+\sqrt{\left(\frac{R_s}{W}\right)^{6}+\frac{1}{2*6^{3}}}\,
\biggr]^{-1/3} \;. \label{cn3}
\end{eqnarray}

For instance, the ground-state energy for cylindric geometry reads
\begin{eqnarray}
e=&& e_{B}+2\pi n_{\lambda}^{-1} \biggr\{ \frac{1}{2}e_{B} \beta 
W (-2 R_s+ \beta W) \rho_{c} \nonumber\\
&&+\left(\frac{1}{2}+\frac{R_s}{W}-\beta\right)\left(\frac{\hbar^2}
{2m}+d_4 \rho_c \right) \rho_c \nonumber\\ 
&&+ \frac{1}{6} W^{2} \rho^2_c \biggr[ b_{4} \left( \frac{1}{4}
+\frac{R_s}{W}-\beta \right) \nonumber\\
&&+\frac{3 c_{4}}{\gamma_4+3} \left( \frac{1}{\gamma_4+4}
+\frac{R_s}{W}-\beta \right) \rho _{c} \biggr] \biggr\} \;.
\label{ecyl}
\end{eqnarray}
Here $\rho_c$ and $R_s$ are functions of $\epsilon$ given by Eqs.\
(\ref{epic1}) and (\ref{radius}), respectively.

\newpage

\begin{table}
\caption{Bulk observables for liquid $^4$He at $T=0$ and the
calculated parameters $\varepsilon_\ell$ and $a_c$.}
\begin{tabular}{lccr}
Observable & Data && Ref. \\
\tableline
$e_B$~[K]                      &    $-7.15$  
                               && \onlinecite{sat2} \\
$\rho_0$~[\AA$^{-3}$]          &    $0.021836$  
                               && \onlinecite{sat2} \\
$\cal{K}$~[K]                  &    $27.2$  
                               && \onlinecite{sat2} \\
$\sigma_{\rm exp}$~[K/\AA$^2$] & $0.274 \pm 0.003$
                               && \onlinecite{gest} \\
                               & $0.257 \pm 0.001$
                               && \onlinecite{isis} \\
                               & $0.272 \pm 0.002$
                               && \onlinecite{raw} \\
 \\
Parameter & Value & Theory & Ref. \\
\tableline
$\varepsilon_\ell$~[K/\AA]     & $1.237$ & Skyrme-DF
                               & PW$^{\rm a}$ \\
$\varepsilon_\ell$~[K/\AA]     & $0.882$ & OP-NLDF   & PW \\
$a_c$~[K]                      & $10.45$ & Skyrme-DF 
                               & \onlinecite{sat1} \\
$a_c$~[K]                      & $10.90$ & & \onlinecite{sat2} \\
$a_c$~[K]                      & $10.86$ & & PW \\
$a_c$~[K]                      & $8.58$ & OP-NLDF & PW \\
$\left(\frac{\pi\,\rho_0}{48}\right)^{1/3}
\frac{a_c}{\varepsilon_\ell}$  & $0.99$ & Skyrme-DF & PW \\
$\left(\frac{\pi\,\rho_0}{48}\right)^{1/3}
\frac{a_c}{\varepsilon_\ell}$  & $1.10$ & OP-NLDF & PW \\
$\left(\frac{\pi\,\rho_0}{48}\right)^{1/3}
\frac{a_c}{\varepsilon_\ell}$  & $1.00$ & DM
                               & Eq.(\ref{curve03})-PW \\

\end{tabular}
$^{\rm a}$PW stands for results obtained in the present work.
\label{table1}
\end{table}

\newpage

\begin{figure}
\caption{Normalized moments $[< r^k(n_\lambda) >]^{1/k}/
n^{1/2}_\lambda$ as a function of the inverse of square root of
longitudinal density $\nu_{\rm cyl} = n^{-1/2}_\lambda$ for free
helium cylinders. Open and full symbols stand for Skyrme-DF and
OP-NLDF results, respectively. The dashed and dot-dashed curves are
only given to guide the eye. The solid lines correspond to the
asymptotic values given by Eq.\ (\protect\ref{radius3}).}
\label{moment1}
\end{figure}

\begin{figure}
\caption{Similar to Fig.\ \protect\ref{moment1}, but for free
helium spheres. The solid lines correspond to the asymptotic values
given by Eq.\ (\protect\ref{radius4}).}
\label{moment2}
\end{figure}

\begin{figure}
\caption{Surface tension calculated by using the Skyrme-DF is shown
as a function of the energy difference $e-e_B$. In order to not
overcrowd the picture only some selected data are plotted and solid
curves are drawn to guide the eye. Dashed lines indicate the linear
asymptotic laws for curved geometries given by Eqs.\
(\protect\ref{sigh1}) and (\protect\ref{sigh2}).}
\label{tension1}
\end{figure}

\begin{figure}
\caption{Same as Fig.\ \protect\ref{tension1} with data calculated
by using the OP-NLDF. In addition, the dashed curve for planar
slabs is the quadratic asymptotic law given by Eq.\
(\protect\ref{sigh01}).}
\label{tension2}
\end{figure}

\begin{figure}
\caption{Thickness of the free surface as a function of the energy
difference $e-e_B$. The circles, triangles, and squares represent
own data for spherical, cylindrical, and planar systems,
respectively. Empty symbols stand for data calculated by using the
Skyrme-DF, while full symbols are results from the OP-NLDF
approach. As in previous figures, only some selected data are
plotted and solid curves are drawn to guide the eye. Diamonds are
OT-NLDF results for spheres from Ref.\ \protect\onlinecite{dali}.
Full and empty five points stars are Monte Carlo simulations for
spheres and slabs reported in Ref.\ \protect\onlinecite{gall}. Six
points stars are Monte Carlo data from Ref.\ \protect\onlinecite{%
vale}. Asterisks are OP-NLDF results for helium adsorbed onto
planar Na and Li substrates. The cross ($\times$), the ``otimes''
($\otimes$), and the ``oplus'' ($\oplus$) are experimental values 
from Refs.\ \protect\onlinecite{sill}, \protect\onlinecite{dali},
and \protect\onlinecite{penal}, respectively.}
\label{width}
\end{figure}

\begin{figure}
\caption{Quantity $W \sigma_A$ as a function of the energy
difference $e-e_B$. As in Fig.\ \protect\ref{width}, the circles,
triangles, and squares represent data for spherical, cylindrical,
and planar systems, respectively. Empty symbols stand for data
calculated by using the Skyrme-DF, while full symbols are results
from the OP-NLDF approach. Dashed lines indicate the linear
asymptotic laws for curved geometries given by Eqs.\
(\protect\ref{prod1}) and (\protect\ref{prod2}).}
\label{prod}
\end{figure}


\begin{references}
\bibitem{abscissa}In this case, results for the thickness are
plotted at the abscissa $\Delta e = e - e_B$ corresponding to free
planar films of $n_c=0.34$~\AA$^{-2}$, but data are slightly
shifted to be distinguished.
\end{references}
\end{document}